\documentclass[graybox, nosecnum]{svmult}

\let\citep=\cite

\usepackage{mathptmx}       
\usepackage{helvet}         
\usepackage{courier}        
\usepackage{type1cm}        

\usepackage{makeidx}         
\usepackage{graphicx}        
\usepackage{multicol}        
\usepackage[bottom]{footmisc}
\usepackage[normalem]{ulem}	
\usepackage{soul}   

\usepackage{amsmath}

\def\be#1{\begin{equation}\label{#1}}
\def\ee{\end{equation}}
\newcommand {\ba}[2]{\be{#1}\begin{array}{#2}}
\newcommand {\ea}{\end{array} \ee}
\def\eq#1{(\ref{#1})}

\newcommand{\qq}{\,,\qquad}
\renewcommand{\=}{\stackrel{\mbox{\scriptsize def}}{=}}
\let\TS=\textstyle
\let\DS=\displaystyle
\def\({\left(}
\def\){\right)}
\let\w = \omega
\def\kB{k_{\!B}}

\def\bb{\vphantom{\Bigl|}}

\def\av#1{\left\langle{#1}\bb\right\rangle}

\newcommand {\Vect} [1] {{\bf #1}}
\newcommand {\Tens} [1] {{\bf #1}}
\def\A{\Vect{A}}
\def\va{\Vect{a}}
\def\F{\Vect{F}}
\def\At{\widetilde{\mbox{\boldmath{$\epsilon$}}}}
\def \Ats{\widetilde{\epsilon}}
\def\e{\Vect{e}}
\def \rr {\Vect{r}}
\def \RR {\Vect{R}}

\def \uu {\Vect{u}}
\def \vv {\Vect{v}}
\def \vt {\tilde{\Vect{v}}}
\def \n  {\Vect{n}}
\def \nab{\stackrel{\circ}{\nabla}}
\def \cst{\mbox{\boldmath{$\tau$}}}

\newcommand{\added}[1]{#1}

\begin{document}

\title*{Discrete and Continuum Thermomechanics}
\author{Anton M. Krivtsov and Vitaly A. Kuzkin}
\institute{Anton M. Krivtsov\at Peter the Great Saint Petersburg Polytechnic University, Polytechnicheskaya st. 29, Saint Petersburg, Russia, \email{akrivtsov@bk.ru}
\and Vitaly A. Kuzkin \at Institute for Problems in Mechanical Engineering RAS, Bolshoy pr. V.O. 61, Saint Petersburg, Russia,
\email{kuzkinva@gmail.com}}
%
%
\maketitle

\begin{abstract}
{}In the present chapter, we discuss an approach for transition from discrete to continuum description of thermomechanical behavior of solids. The transition  is carried out for several anharmonic systems: one-dimensional crystal, quasi-one-di\-men\-sio\-nal crystal (a chain possessing longitudinal and transversal motions), two- and tree-dimensional crystals with simple lattice.  Macroscopic balance equations are derived from equations of motion for particles.  Macroscopic parameters, such as stress, heat flux, deformation, thermal energy, etc., are represented via parameters of the discrete system. Closed form equations of state relating thermal pressure, thermal energy and specific volume are derived. Description of the heat transfer in harmonic approximation is discussed. Unsteady ballistic heat transfer in a harmonic one-dimensional crystal is considered.  The heat transfer equation for this system is rigorously derived.
\end{abstract}

\section{Synonyms}
Discrete and continuum thermoelasticity; Thermoelasticity: from particle dynamics to continuum mechanics; Mechanics of discrete media: thermoelasticity



\section{Introduction}

The general laws used to construct equations of continuum mechanics leave too much freedom for choosing the constitutive equations~\citep{Zhilin 2013}. Therefore discrete mechanical models can add a lot for understanding of the general nature of thermomechanical processes in solids~\citep{Hoover 2013, Lepri 2016, Krivtsov 2007 book, Weinberger 2016}. Lattice models of solids are good examples of discrete systems where rigorous analytical solutions can be constructed. Recent advances in nanotechnologies have brought these lattice models much closer to reality, showing that solids with perfect lattices can be indeed realized at least at nanoscale. In the nearest future this could be done for much higher scale levels providing high quality materials with extraordinary thermomechanical properties.

\medskip
\noindent
Passage from discrete to continuum mechanics requires a number of important steps. Below we outline some of them, which in many cases allow to rigorously obtain macroscopic equations of continuum mechanics from microscopic equations of discrete mechanics.

\begin{description}

\item[\bf Averaging.] The important feature of discrete mechanics is the existence of the chaotic thermal motion of particles forming the material. In continuum mechanics, it is taken into account implicitly via constitutive relations. Therefore some averaging procedure should be used in order to bridge the gap between discrete and continuum descriptions. Averaging can be carried out over space, time, or over a set of random realizations of the considered mechanical system. The latter averaging is preferable for both analytical derivations and computer simulations. In rigorous derivations, this average can be replaced by mathematical expectation.

\item[\bf Power expansion.] In general, averaging can not give closed equations for anharmonic crystals. However, in the case of small nonlinearity, series expansion yields  closed system of thermomechanical equations.

\item[\bf Similarity approximation.] Additional relations between different terms in the power expansion can be obtained using the similarity approximation. The approximation states that averages of higher powers are represented in terms of the lower ones~(see e.g. formula~\eq{17}).

\item[\bf Correlational analysis.] Obtaining closed equations for heat transfer processes requires averaging the quantities defined for all pairs of particles. The quantities characterize correlation of motion characteristics~(e.g. velocities) at different spatial points. This approach yields an exact analytical description of  the heat transfer in harmonic crystals.

\item[\bf Motion separation.] In discrete systems containing large number of particles, the different time scale apply for the different processes. Local transitional processes are realized at the time scales of the individual particles oscillations. These processes result in thermodynamically equilibrium states such as local energy equipartition. The nonlocal processes such as propagation of mechanical waves and heat transfer are much slower. Therefore significant simplification can be achieved if these processes are separated.

\item[\bf Continualization.] The macroscopic quantities are supposed to be varying slowly in space  at distances of order of particles separation. Then series expansion replacing finite differences with spatial derivatives can be used.

\end{description}

\noindent
Summarizing the above, stochastic finite difference equations at microscale are to be replaced by deterministic partial differential equations at macroscale.

The transition from discrete to continuum equations is carried out below for several simple anharmonic systems: one-dimensional crystal, quasi-one-di\-men\-sio\-nal crystal (a chain possessing longitudinal and transversal motions), two- and tree-dimensional crystals with simple lattice. The anharmonicity is essential to observe  coupled thermomechanical phenomena such as thermal expansion and etc. For all these systems, thermomechanical equations are derived in the adiabatic approximation~(heat transfer is neglected). Rational description of the heat transfer processes is much more complicated. The complexity is caused by anomalous nature of the heat transfer in perfect crystalline solids. Such processes are considered in the last part, where the heat transfer in the one-dimensional crystal is considered in harmonic approximation and equations of unsteady heat processes are rigorously derived.

\medskip

\section{Basic Designations and Assumptions}

The discrete system is a set of particles interacting via prescribed interparticle forces of a potential nature. The particles are arranged in a perfect crystal lattice and they perform permanent random oscillations (thermal motion) in the vicinity of the lattice nodes. The nodes participate in much more smooth mechanical motion that can be continualized.

A set of different realizations of the same system is considered. The realizations differ only by the random component of the motion. Let $\phi_k^{(s)}$ be the realization number $s$ of physical property~$\phi$ for the particle number $k$. Then the mathematical expectation~$\overline{\phi}_k$ for the random value $\phi_k^{(s)}$ is defined as
\be{1}
    \overline{\phi}_k \= \av{\phi_k^{(s)}} \= \lim_{S\to\infty}\frac1S\sum_{s=1}^S \phi_k^{(s)}.
\ee
In numerical simulations, the mathematical expectation is replaced by an average over a sufficiently large number of realizations~$S$. Alternatively the averages over space and time can be used. However definition~\eq{1} is preferable, since it contains no additional parameters such as averaging period or averaging volume. Additionally, in the case of computer simulations, formula~\eq{1} requires less computations.

Any physical quantity can be represented as a sum of the regular component \eq{1} and a stochastic component:
\be{3}
    \phi_k^{(s)} = \overline{\phi}_k + \widetilde{\phi}_k^{(s)}
    ,\qquad\mbox{where}\qquad \widetilde{\phi}_k^{(s)}\=\phi_k^{(s)} - \overline{\phi}_k
.\ee
Index ${}^{(s)}$ is omitted below for brevity.

The mathematical expectation is supposed to be slowly varying in space. Therefore continualization with respect to spatial coordinate can be used. For one-dimensional case, the continualization is carried out as follows
\be{2}
    \overline{\phi}_{k} \= \phi(x) \qq
    \overline{\phi}_{k+1} \= \phi(x+a) = \phi(x) + a\phi'(x) 
    + O(a^2)
,\ee
where $x$ is the continuous spatial variable equal to $ka$ at the lattice nodes, $O(\varepsilon)$ is a quantity of the same order as $\varepsilon$.

To describe temperature of the crystal at microlevel the kinetic temperature
can be used, which is defined as
\be{k1}
    \frac d2\,\kB T
    = K_T
\qq
    K_T\= \av{\frac {m {\tilde v}^2}2}
,\ee
where $\kB$ is the Boltzmann constant,
$d=1,2,3$ is the space dimension,
$K_T$ is the expected value of thermal kinetic energy of the particle,
$m$ and $\tilde v$ are the mass and the stochastic velocity of the particle. \added{The space dimensionality, $d$, in formula~\eq{k1} is introduced in order to guarantee that at  equilibrium kinetic energy per degree of freedom is equal to~$\frac{k_B T}{2}$~(the equipartition theorem, see e.g.~\citep{Hoover 1991 book}).}

Formula \eq{k1} allows to calculate temperature for any particle at any given moment of time.
This definition of temperature can be proved by considering an ideal gas as a thermometer (Hoover et al. 1993). Such thermometer being connected to the crystal after equilibration shows exactly the same temperature, as the kinetic temperature of the crystal~\eq{k1}.

\section{One-dimensional Crystal}

\subsection{General Equations}

Consider one-dimensional crystal with nearest-neighbor interactions via potential $\Pi(r)$, where $r$ is the interparticle distance. The dynamics equations for the longitudinal motions are
\be{4}
    m\ddot u_k = F_{k+1}-F_k \qq
    F_k \= -f(a+\epsilon_k) \qq
    \epsilon_k \= u_{k}-u_{k-1}.
    \ee
Here
 $m$ is the particle mass;
 $u_k$ is the displacement of the particle number $k$; \added{$F_k$ is the force acting on particle~$k-1$ from particle~$k$}; $f(r)\=-\Pi'(r)$;
 $\epsilon_k$ is the bond deformation;
 $a$ is the lattice step.

Exact transformations of equations~\eq{4} yields the following energy balance equation~\citep{Krivtsov 2007 book}:
\be{5}
    \dot{\cal E}_k = {\cal H}_{k}-{\cal H}_{k+1},
\ee
where ${\cal E}_k$ is the specific energy and ${\cal H}_k$ is the energy flux, defined as
\be{6}
    2{\cal E}_k \= m v_k^2 + \Pi(a+\epsilon_{k+1}) + \Pi(a+\epsilon _{k})
\qq
    2{\cal H}_k \= -F_k(v_k+v_{k-1}),
\ee
where $v_k\=\dot u_k$ is the particle velocity.

Calculating mathematical expectation~\eq{1} and performing continualization~\eq{2} of microscopic balance equations~\eq{4}--\eq{5} yields macroscopic balance equations
\be{7}
    m\rho_0\dot v = p' \qq
    \rho_0 \dot U = -p v' - h',
\ee
where terms $O(a^2)$ are neglected,
prime stands for derivative with respect to spatial variable $x$,
\be{8a}
    \rho_0\=1/a
\qq
    V\=a + au'
\ee
 \added{are the number of particles in a volume~(the reverse specific volume) for the reference configuration and the actual specific volume,}
$u = \overline u_k$
and
$v = \overline v_k$
are the macroscopic displacement and velocity,
\be{8}
    p \= \av{f(V+{\widetilde\epsilon}_k)}
\qq
    U \= \frac m2\av{{\widetilde v}_k^2}
    + \av{\Pi(V+{\widetilde\epsilon}_k)}
\qq
    h \= -\av{{\widetilde F}_k{\widetilde v}_{k-1}}
\ee
are the pressure, the specific internal energy (per particle), and the heat flux. \added{Note that in the one-dimensional case volume~$V$ has the same dimension as length.}

To close system~\eq{7}, the constitutive equations for~$p$, $U$ and $h$ are required.
Application of the virial theorem, which is a particular case of the mentioned above motion separation, gives the following representation for the internal energy~\citep{Krivtsov 2007 book}:
\be{9}
    U = -\frac 12\av{f(V+{\widetilde\epsilon}_k){\widetilde \epsilon}_{k}}
    + \av{\Pi(V+{\widetilde\epsilon}_k)}
.\ee

It is convenient to represent the pressure and internal energy as sum of the cold and thermal terms:
\be{10}
    p = p_0 + p_T \qq
    U = U_0 + U_T
,\ee
where
\be{11}
    p_0 \= p|_{\tilde\epsilon_k=0} = f(V) \qq
    U_0 \= U|_{\tilde\epsilon_k=0} = \Pi(V)
\ee
are the cold pressure and energy (corresponding to absence of the thermal motion),
$p_T$ and $U_T$ are the thermal pressure and energy.
Using these quantities equations \eq{7} can be represented in the from
\be{11a}
    m\rho_0\dot v = p_0' + p_T' \qq
    \rho_0 \dot U_T = -p_T v' - h'
.\ee
\added{Here $U_T$ is the specific thermal energy per particle, $m\rho_0=m/a$ is the mass density.}

\subsection{The First Thermal Approximation}

According to formulas \eq{8}--\eq{11} the thermal pressure and the thermal energy depend on a single microscopic parameter --- the stochastic part of deformation ${\widetilde \epsilon}_{k}$. Assuming this parameter to be small the power expansion can be used for derivation of constitutive equations.

The first nontrivial approximation for equations \eq{8} and \eq{9}
in terms of the small parameter ${\widetilde \epsilon}_{k}$
yields
\be{12}
    p_T \approx f_2(V)\xi \qq
    U_T \approx f_1(V)\xi,
\ee
where
\be{13}
    \xi \= \av{{\widetilde\epsilon}_k^2}
\qq
    f_p(V) \= \frac{(-1)^p}{p!}\,\frac{{\rm d}^p}{{\rm d}V^p}\,f(V)
.\ee
Parameter $\xi$ can be refereed to as the deformation temperature. Elimination of this quantity from~\eq{12} yields the Mie-Gr\"uneisen equation
of \added{state~\citep{Mie 1903, Gruneisen 1912}. The equation gives linear relation between thermal pressure and thermal energy:}
\be{14}
    p_T = \Gamma(V)\,\frac{U_T}V \qq
    \Gamma(V) = V\,\frac{f_2(V)}{f_1(V)}
,\ee
where $\Gamma(V)$ is dimensionless Gr\"uneisen parameter.

Substitution of expressions \eq{8a}, \eq{10}, \eq{11}, \eq{14} into the microscopic balance equations \eq{11a} yields two equations for three macroscopic variables: displacement $u$, thermal energy~$U_T$, and heat flux $h$. If the heat flux could be expressed in the terms of $u$ and~$U_T$ (similarly to the pressure and internal energy) then the closed system of macroscopic equations would be obtained. However, in general this is impossible. The simplest way to close the system is to apply the adiabatic approximation, which is $h\equiv0$. This approximation is good for relatively fast processes, such as shock waves.

For the adiabatic approximation the following nonlinear wave equation can be obtained~\citep{Krivtsov 2007 book}:
\be{15}
    \ddot u - c^2 u'' = 0 \qq
    c^2 \= \frac1{m\rho^2} \(f_1 + (3f_1f_3-f_2^2)f_1^{-2}U_T\bb\),
\ee
where $f_p=f_p(V)$ and thermal energy $U_T$ satisfies relation
\be{16}
    U_T = B \sqrt{f_1(V)}
    \qq V = a+au'
.\ee
Constant $B$ can be determined from the initial conditions. In the case of small deformations, equation \eq{15} is a linear wave equation where the sound speed $c$ depends on the thermal energy.

\subsection{The Second Thermal Approximation}

The Mie-Gr\"uneisen equation of state \eq{14} is not valid \added{when $f_1$ is close to zero, while $f_2$ is nonzero. This case corresponds to strong tension of the crystal up to the breakage point. Then it is essential to leave higher order terms in series expansion of the thermal energy and thermal pressure with respect to~${\widetilde\epsilon}_k$. To close the resulting system of equations, the similarity assumption~\citep{Krivtsov 2003 CSF} is used:}
\be{17}
    \av{\widetilde{\epsilon}_k^4}
    = \lambda \av{\widetilde{\epsilon}_k^2}^2
    = \lambda \xi^2
,\ee
where $\lambda$ is an empirical constant. At small temperatures~$\lambda \approx 3$~\citep{Krivtsov 2007 book}. Then constitutive equations~\eq{12} take the form
\be{18}
    p_T \approx f_2(V)\xi \qq
    U_T \approx f_1(V)\xi + \frac34\lambda f_3(V)\xi^2
.\ee
Elimination of $\xi$ yields the following extension of the Mie-Gr\"uneisen equation of state \eq{14}:
\be{19}
    p_T = 2f_2(V)\,\frac{\sqrt{f_1^2(V)+3\lambda f_3(V)U_T}-f_1(V)}{3\lambda f_3(V)}
.\ee
In the vicinity of the critical tension (relevant to the bond break) the term $f_1(V)$ is close to zero and equation \eq{15} takes the form
\be{20}
    p_T
    = 2f_2(V)\,\sqrt\frac{U_T}{3\lambda f_3(V)}
.\ee
Thus in this case the thermal pressure is proportional to the square root of the thermal energy that is substantially different from the Mie-Gr\"uneisen equation of state, which predicts a linear dependence.

The wave equation~\eq{15} shows that the sound speed in the hot crystal vanishes in the point of the ``hot break'', which is located shortly before  the point of the ``cold break''~$V=V_1$, where $f_1(V_1)=0$. The equation of state in the vicinity of the point of the ``hot break'', $V=V_1-\zeta$, is represented in the following form~\citep{Krivtsov 2007 book}:
\be{21}
    p_T(V_1-\zeta,U_T)
    = \frac{2f_2(V_1)}{\sqrt[3]{f_2(V_1)}}\,U_T^{\frac23}
\qq
    \zeta \=
    \sqrt[3]{\frac{U_T}{8f_2(V_1)}}
.\ee
This asymptotics has power $2/3$ rather then $1/2$ that holds in representation \eq{20}. The reason for this difference is that the location of the ``hot break'' depends on the thermal energy itself. Note that power~$2/3$ inherent to the ``hot break'' is between~$1$ (the Mie-Gr\"uneisen equation of state) and $1/2$ (``cold break'').

More accurate equations of state can be obtained leaving higher order terms in the expansion~\eq{18}. For example, adding the fourth order term to the expansion of $p_T$ yields an appropriate description of weak anharmonic effects for any tension (not only in the vicinity of the break point). Further expansions leaving higher powers of $\xi$ require the following generalization of the similarity assumption \eq{17}:
\be{22}
    \av{\widetilde{\epsilon}_k^{2p}}
    = \lambda_p \av{\widetilde{\epsilon}_k^2}^p
    = \lambda_p \xi^p
,\ee
where $\lambda_p$ are empirical constants. In principle, an infinite expansion describing
arbitrary nonlinearity can be carried out.
However numerical simulations show that formulas~\eq{22} are less accurate for higher~$p$~(parameters~$\lambda_p$
depend on thermal energy and deformation of the chain).

\subsection{Kinetic Temperature}

Definition of kinetic temperature \eq{k1} for one-dimensional case ($d=1$) together with equations \eq{8}--\eq{9} yields
\be{23}
    U = \frac12\,\kB T
    + \av{\Pi(V+{\widetilde\epsilon}_k)}
\qq
    \kB T = -\av{f(V+{\widetilde\epsilon}_k){\widetilde \epsilon}_{k}}
.\ee
These formulas allow rewriting the above equations of state in  terms of temperature instead of the thermal energy. In particular, the first thermal approximation \eq{18} yields
\be{24}
    p_T
    = \frac{f_2(V)}{f_1(V)}\,\kB T
\qq
    U_T
    = \kB T
,\ee
where $\kB T = f_1(V)\xi$.
In this case, as it follows from the virial theorem, kinetic and potential parts of the thermal energy are equal, the heat capacity (for the constant volume) is constant and equal to the Boltzmann constant $\kB$.
Then the equation of state reads
\be{25}
    p = f(V) + \frac{f_2(V)}{f_1(V)}\,\kB T.
\ee
For the case of small deformations~$|u'|\ll1$, substitution~$V=a+au'$ to
equation~\eq{25} yields
\be{26}
    \tau = f_1(a)\,\epsilon - \frac{f_2(a)}{f_1(a)}\,\kB T
\qq
    \tau \= f(a) - p
,\ee
where $\tau$ is the small stress, $\epsilon = u'$ is the small strain. Equation \eq{26} is a one-dimensional version of the Duhamel--Neumann law of linear thermoelasticity.
From equation~\eq{26} the coefficient of the thermal expansion is calculated as
 \be{27}
    \alpha
    \= \left.\frac{\epsilon}{T}\right|_{\tau=0}
    = \kB\,\frac{f_2(a)}{{f_1}^{\!2}(a)}
.\ee

If nonlinear temperature effects are taken into account, the kinetic and potential parts of the thermal energy are not equal, and the thermal energy is no longer proportional to temperature. An explicit representation of the thermal energy in terms of the temperature is obtained using series expansions in equations~\eq{23}. The second approximation yields
\be{28}
    U_T \approx f_1(V)\xi + \frac34\lambda f_3(V)\xi^2
\qq
    \kB T \approx f_1(V)\xi + \lambda f_3(V)\xi^2
,\ee
Then  internal energy and temperature are related as
\be{29}
    U_T = \kB T - \frac{\lambda f_3(V)}{4{f_1}^{\!2}(V)}\,(\kB T)^2
,\ee
where terms of the third and higher orders are omitted. \added{Note that the second coefficient in formula~\eq{29} depends on the fourth derivative of the interatomic potential. This fact can be used for calibration of parameters of the potential using experimental  data on the temperature dependence of heat capacity.}

Formula~\eq{29}  is not valid in the vicinity of the break point, where $f_1$ tends to zero. In this case, from \eq{28} it follows that
\be{30}
    U_T = \frac34\,\kB T.
\ee
Note that coefficients of linear terms in formulas~\eq{29}, \eq{30} are different~($1$ vs. $3/4$).

In order to get nonlinear corrections to formula~\eq{30}, additional terms in expansions~\eq{28} are required.

\section{Quasi-one-dimensional Crystal}
\subsection{General Equations}
In the present section, longitudinal and transverse vibrations of an infinite chain with nearest-neighbor interactions are considered~\citep{Kuzkin 2015 PSS}. Each particle has two degrees of freedom. The chain is preliminary stretched. Continualization of this system can be carried out as described in the previous section. The present section, focuses on derivation of equations of state closing the balance equations in the adiabatic approximation.

The steady state of the system is considered. In this case, mathematical expectations of characteristics associated with all particles are identical. Vector connecting two neighboring particles is represented as a sum of its mathematical expectation~$\A$ and the stochastic part~$\At$ such that~$\av{\At} = 0$.
Particles interact via pair potential~$\Pi$.

Consider derivation of equations of state relating pressure, thermal energy
and deformation of the chain. Pressure is defined as
\be{pressure_def}
 p = -\av{\F}\cdot\e, \quad \F = -\Phi\(|\A + \At|^2\) \(\A + \At\), \quad \Phi(A^2) =  -\frac{\Pi'(A)}{A},
\ee
where~$A=|\A|$, $\e$ is the unit vector directed along the chain, $\F$ is the force acting between two neighboring particles.
The pressure is represented as a sum of cold and thermal parts~\eq{10}:
\be{}
  p = p_0 + p_T,
  \qquad
  p_0 = p|_{\At=0} = \Phi\(A^2\),
  \qquad
  p_T = p - p_0.
\ee
The specific energy has the form
\be{therm_en_def}
\begin{array}{l}
  \DS U = \frac{m}{2}\av{\vt^2} + \av{\Pi\(|\A + \At|\)},
  \quad
  U_0 = U|_{\At=0} = \Pi\(A\),
  \quad
  U_T = U-U_0.
\end{array}
\ee
The kinetic part of specific energy is represented as a function of
vector~$\At$ using the Virial transformation~\citep{Krivtsov 2003 CSF}:
\be{KT_virial}
 \frac{m}{2}\av{\vt^2} \approx \frac{1}{2}\av{\At \cdot \F\(\A + \At\)}.
\ee
Formulas~\eq{pressure_def}, \eq{therm_en_def}, \eq{KT_virial} show that the thermal pressure~$p_T$ and thermal energy~$U_T$ depend on vector~$\At$.

\subsection{The First Thermal Approximation}
In order to derive equations of state, consider series expansion of the thermal pressure and thermal energy with respect to~$\At$.
The first approximation yields
\be{pU1Dq}
\begin{array}{l}
\DS p_T \approx \(\Phi'A(\Tens{E} + 2\e\e) + 2\Phi'' A^3\e\e\) \cdot\cdot \av{\At\At},\\[4mm]
\DS U_T \approx -\(\frac{1}{2}\Phi\Tens{E} + \Phi'A^2\e\e\)  \cdot\cdot \av{\At\At},
\end{array}
\ee
where~$\Tens{E}$ is the unity tensor; argument~$A^2$ of functions~$\Phi^{(n)}$ is omitted for brevity; \added{double dot product of tensors $\Vect{a}\Vect{b}$  and $\Vect{c}\Vect{d}$ is defined as~$\Vect{a}\Vect{b}\cdot\cdot\Vect{c}\Vect{d} = (\Vect{a}\cdot\Vect{d})(\Vect{b}\cdot\Vect{c})$}.
Then thermal energy and thermal pressure are proportional
to tensor~$\av{\At\At}$. The tensor characterizes longitudinal and transverse deformations of the bonds caused by thermal motion:
\be{AalAal}
\begin{array}{l}
\DS   \At = \Ats_l \e + \Ats_{t} \n,
\qquad
 \DS \av{\At\At}= \av{\Ats_l^2}\e\e + \av{\Ats_t^2}\n\n + \av{\Ats_l\Ats_t}\(\e\n + \n\e\).
\end{array}
\ee
From formulas~\eq{pU1Dq}, \eq{AalAal} it follows that  thermal pressure and thermal energy depend on~$\av{\Ats_l^2}$, $\av{\Ats_t^2}$. Therefore  system~\eq{pU1Dq} is not closed and additional relation between these parameters is required.
The relation is derived using the equipartition theorem~\citep{Hoover 1991 book}.
\added{The theorem states that kinetic energy is equally distributed between degrees of freedom. In particular, kinetic energies corresponding to longitudinal and transverse motions of the quasi-one-dimensional chain are equal.
Using this fact, the following relation is derived~\citep{Kuzkin 2015 PSS}}:
\be{equip_lin}
 \Pi'' \av{\Ats_l^2} \approx \frac{\Pi'}{A}\av{\Ats_t^2}.
\ee
Covariance of longitudinal and transverse deformations~$\av{\Ats_l\Ats_t}$ do not contribute to the equation of state.

Excluding~$\av{\Ats_l^2}$ and $\av{\Ats_t^2}$ form
the expressions for pressure and thermal energy and using
formula~\eq{equip_lin} yields the equation of state
in the Mie-Gr\"{u}neisen form:
\be{Gamma}
\begin{array}{l}
 \DS p_T = \frac{\Gamma(A)}{A}U_T,
 \quad
 \Gamma  =  \frac{\Gamma_{l}+\Gamma_{t}}{2},
 \quad
 \Gamma_{l} =  -\frac{\Pi''' A}{2\Pi''},
 \quad
\DS \Gamma_{t} = -\frac{\Pi'' A - \Pi'}{2\Pi'}.
\end{array}
\ee
Gr\"{u}neisen parameter~$\Gamma$ has two contributions~$\Gamma_{l}$, $\Gamma_{t}$ from longitudinal and transverse vibrations respectively. The contribution of longitudinal vibrations~$\Gamma_{l}$ is caused by anharmonicity of the interatomic potential only. It vanishes in the case of harmonic potential~($\Pi'''=0$). In contrast, the contribution of transverse vibrations~$\Gamma_{t}$ does not depend on anharmonic properties of the potential. It is caused by geometrical nonlinearity.

\subsection{The Second Thermal Approximation}
The Mie-Gr\"uneisen equation of state~\eq{Gamma} is inapplicable, for example, in the
case of small deformations of the chain~\citep{Kuzkin 2015 PSS}. More accurate equations of state are derived as follows. Thermal pressure and thermal energy are expanded into series with respect to parameter~$\At$ up to the terms of the fourth order. It is assumed that transverse deformations of the bonds are larger than longitudinal~$\av{\Ats_t^2} \gg \av{\Ats_l^2}$.  In order to close the resulting system of equations, the following similarity assumptions
are used:
\be{mu_lambda}
\av{\Ats_t^4} = \lambda \av{\Ats_t^2}^2, \qquad A\av{\Ats_l\Ats_t^2} = \mu \av{\Ats_t^2}^2.
\ee
\added{The second formula in~\eq{mu_lambda} is written assuming that~$\Ats_l$ has the same order as~$\Ats_t^2$.}
Parameters~$\lambda$ and $\mu$  are estimated
using computer simulations. For example, in the case of the Lennard-Jones
potential~$\lambda \approx 3$, $\mu \approx -1$.
Then equation of state similar to equation~\eq{19} is obtained
\be{EOS_ex2}
   \DS p_T = \frac{B_2}{B_4}U_T + \frac{(B_2B_3-B_1B_4)(B_3 -\sqrt{B_3^2+4B_4U_T})}{2B_4^2},
\ee
where
\be{EOS_im3}
  \begin{array}{l}
   \DS B_1 = \Phi'A + \frac{\Gamma_l}{A}\Phi,
  \quad
  B_3\!=\!-2\Phi\!,
  \quad B_4\!=\!-\frac{7}{4}(\lambda\!+\!2\mu)\Phi'\!,\\[4mm]
   \DS B_2= \frac{\mu -\Gamma_l(\lambda+\mu)}{A}\Phi' + \frac{\lambda+4\mu}{2}\Phi''A.
  \end{array}
\ee
Analysis of equation of state~\eq{EOS_ex2} shows that the dependence of  thermal pressure on  thermal energy is strongly nonlinear. For example, consider series expansions of~$p_T(U_T)$ for the cases
of stretched chain~($A > a$), unstretched chain~($A = a$),
and deformation corresponding to zero Gr\"{u}neisen parameter~($\Gamma(A_{*})=0$). Then
\be{EOS_poly}
\begin{array}{l}
   \DS p_T \approx \frac{\Gamma}{A}U_T + \frac{B_2A-\Gamma B_4}{4\Phi^2A} U_T^2,\quad A > a,
\\[4mm]
\DS p_T \approx -\sqrt{\frac{2\Pi''\,U_T}{7(\lambda+2\mu)}},\quad A=a,
\\[4mm]
\DS p_T \approx \frac{B_2}{4\Phi^2} U_T^2, \quad A=A_{*}.
\end{array}
\ee
Formulas~\eq{EOS_poly} and results of molecular dynamics  simulations~\citep{Kuzkin 2015 PSS} show that in the case of small deformations the Mie-Gr\"uneisen equation of state is inaccurate. Moreover in the case of unstretched chain or deformation corresponding to zero Gr\"uneisen parameter, the Mie-Gr\"uneisen equation is inapplicable. In these cases, nonlinear equation of state~\eq{EOS_ex2} should be used.

Thus the approach described above allows to derive nonlinear equations of state. The equations are applicable in wider range of deformations and thermal energies than the Mie-Gr\"uneisen equation.

\section{Two- and Three-dimensional Crystals}
\subsection{General Equations}
In the present section, continuum balance equations and equations of state are derived from lattice dynamics equations for two- and three-dimensional crystals with simple structure~\citep{Krivtsov 2007 book, Krivtsov 2011 MTT, Kuzkin 2010 IUTAM, Kuzkin 2015 PhysMesomech}.

Consider an infinite crystal lattice with simple structure in $d$-dimensional
space~($d=1, 2$ or $3$). Two states of a crystal and its equivalent continuum are considered: the reference configuration~(undeformed crystal) and the current configuration. Thermoelastic deformations of the crystal are investigated. In this case, the mapping between the reference and current  configurations exists. Radius-vectors of equivalent continuum in the reference and current configurations are denoted as~$\rr$ and $\RR$ respectively.

Relations between continuum deformation measures and deformations of bonds in a crystal are derived as follows.
Consider a reference particle. Neighbors of the reference particle are marked by index~$\alpha$. Vector connecting the particle with its neighbor number~$\alpha$ in the reference configuration is denoted as~$\va_\alpha$. By the definition
vectors~$\va_\alpha$ have the following property
\begin{equation}\label{kuzkin:eq:vavma}
   \va_\alpha = -\va_{-\alpha}.
\end{equation}
In the current configuration, vector connecting the reference particle and it's neighbor~$\alpha$ is represented as a sum of its mathematical expectation~$\A_\alpha$ and the remaining oscillatory part~$\At_\alpha$ such that~$\av{\At_\alpha} = 0$.
It is assumed that mathematical expectations of particle positions are
identical to positions of corresponding points of continuum. Then continualization
of vector~$\A_\alpha$ yields
\be{Aa}
\begin{array}{l}
\DS \A_\alpha = \RR(\rr+\va_\alpha) - \RR(\va_\alpha)
\approx \va_\alpha \cdot \nab\RR,
\end{array}
\ee
where $\nab$ is the nabla operator in the reference configuration.
In the literature,  formula~\eq{Aa} is refereed to as the Cauchy-Born rule.

The expression for the strain gradient follows from formula~\eq{Aa}:
\be{str grad}
\nab \RR = \(\sum_{\alpha} \va_\alpha\va_\alpha\)^{-1}
 \cdot \sum_{\va} \va_\alpha \A_\alpha.
\ee
Thus formulas~\eq{Aa}, \eq{str grad} relate deformations of the bonds in a crystal with deformations of the equivalent continuum.

Consider derivation of  continuum balance equations from discrete equations of motion of the crystal.
Equation of motion for the reference particle reads
\be{EM}
  m \ddot\uu  =  \sum_\alpha \F_\alpha,
\ee
where ${{\bf F}_\alpha}$ is the force acting on the reference particle from its neighbor~$\alpha$, $m$ is particle's mass. Mathematical expectation of both parts of equation~\eq{EM}  is calculated. Continualization is carried out using the third Newton's law
\be{}
 \F_\alpha(\rr) = -\F_{-\alpha}(\rr + \va_\alpha), \qquad  \av{\F_\alpha}(\rr) \approx -\av{\F_{-\alpha}}(\rr)
 -
 \va_\alpha \cdot\nab  \av{\F_{-\alpha}}(\rr).
\ee
Then equation of motion~\eq{EM} in continuum limit takes the form
\be{EMcl}
 \frac{m}{V_0} \av{\ddot{\uu}} = \nab \cdot
 \( \frac{1}{2V_0}\sum_\alpha \va_\alpha \av{\F_\alpha}\),
 \qquad
 V_0 = \frac{\sqrt{5-d}}{2} a^d,
\ee
where $a$ is an equilibrium distance, \added{$V_0$ is the volume of the
elementary cell in the reference configuration~(volume per particle in an
infinite perfect lattice~\citep{Krivtsov 2007 book})}.
%
%

Equation~\eq{EMcl} has the same form as continuum momentum balance equation in the reference configuration. Comparison of these equations yields the expression for
 the Piola stress tensor~${\bf P}$:
\be{Piola stress}
  {\bf P} = \frac{1}{2V_0}\sum_\alpha {\bf a}_\alpha \Bigl\langle{{{\bf F}_\alpha}}\Bigr\rangle.
\end{equation}
Similar derivations in the current configuration
yields the expressions for the Cauchy stress tensor:
\be{Cst}
\cst =
 \frac{1}{2V}\sum_\alpha \A_\alpha \av{\F_\alpha},
\end{equation}
where~$V$ is the volume per particle in the current configuration.
Formulas~\eq{Piola stress}, \eq{Cst} represent Cauchy and Piola stress tensors via interatomic forces and distances. Note that this approach allows to calculate the stress field to
the accuracy of tensor with zero divergency.

Consider the equation of energy balance for the reference particle~\citep{Kuzkin 2010 IUTAM}. Body forces and volumetrical heat sources are neglected. Derivations are carried out in the reference configuration.
Specific energy per particle has the following form
\begin{equation}\label{Utdef}
 U  =   \frac{m}{2} \av{\vt^2}
 +
 \frac{1}{2}\sum_\alpha \av{\Pi(|\A_\alpha + \At_\alpha|)}.
\end{equation}
Calculating time derivative and performing continualization yields
\be{EEB}
  \frac{\dot{U}}{V_0} = {\bf P} \cdot\cdot \(\nab\av{\vv}\)^T
  +
  \nab \cdot \(\frac{1}{2V_0}\sum_\alpha \va_\alpha
\av{ \widetilde{\F}_\alpha \cdot \vt}\).
\ee
Comparison of equation~\eq{EEB} with continuum equation of energy balance yields the expression for the heat flux in the reference configuration:
\be{hfrc}
 \Vect{h} = -\frac{1}{2V_0}\sum_\alpha \va_\alpha
\av{\F_\alpha \cdot \vt}.
\ee
Equivalent expressions for the heat flux are the following
\be{hfrc2}
\begin{array}{l}
\displaystyle {\bf h}
=
-\frac{1}{4V_0}\sum_\alpha \va_\alpha \av{\F_\alpha \cdot
(\vt+\vt_\alpha)}
=
-\frac{1}{2V_0}\sum_\alpha \va_\alpha \av{\F_\alpha\cdot \vt_\alpha}.
  \end{array}
\end{equation}
Expressions~\eq{hfrc}, \eq{hfrc2} coincide in the continuum limit. Note that according to formulae~\eq{EEB}, the heat flux is calculated to the accuracy of vector field with zero divergency.

Expressions for the heat flux in the current configuration are obtained using
the identity~${\bf H} = V_0\(\nab\RR\)^T \cdot {\bf h}/V$:
\be{hfcc}
\begin{array}{l}
\DS  {\bf H}= -\frac{1}{4V}\sum_\alpha \A_\alpha
\av{\F_\alpha\cdot(\vt+\vt_\alpha)}
 =
 -\frac{1}{2V}\sum_\alpha
\A_\alpha \av{\F_\alpha\cdot\vt_\alpha}
= -\frac{1}{2V}\sum_\alpha \A_\alpha
\av{\F_\alpha \cdot \vt}.
  \end{array}
\end{equation}

Thus continuum balance equations are derived from particle dynamics equations. The expressions for equivalent stress tensors and heat fluxes are given by formulas~\eq{Cst}, \eq{hfcc}. In the following section, the expression for the Cauchy stress tensor is used for derivation of equations of state.

Derivations presented above are based on the assumption that the total force acting on a particle is represented as a sum of forces~$\F_\alpha$. In the case of pair interactions, this assumption is satisfied identically. The case of multibody interactions is discussed in paper~\citep{Kuzkin 2010 PRE}. It is shown that similar decomposition of the total force can be carried out in the case of an arbitrary multibody potential. Therefore formulas~\eq{Cst}, \eq{hfcc} are valid in the case of multibody interactions.

\subsection{The First Thermal Approximation}
In order to close balance equations described in the previous section, additional constitutive relations are required. Consider the equation of state for the stress tensor.

The stress tensor is represented as a sum of
cold and thermal parts:
\be{}
 \cst = \cst_0  + \cst_T, \qquad \cst_0= \cst|_{\At_\alpha=0} = -\frac{1}{2V}\sum_\alpha \Phi(A_\alpha^2) \A_\alpha\A_\alpha,
 \qquad
 \cst_T = \cst - \cst_0,
\ee
where function~$\Phi$ is defined by formula~\eq{pressure_def}. Cold stresses are represented as a function of deformation measure using formulas~\eq{Aa}, \eq{Cst}~(see paper~\citep{Krivtsov 1999 ZAMM}):
\be{}
 \cst_0 = -\frac{1}{2V_0 \sqrt{\det\Tens{C}}} (\nab \RR)^T \cdot \sum_{\alpha} \Phi(\va_\alpha\va_\alpha \cdot\cdot
 \Tens{C}) \va_\alpha\va_\alpha \cdot \nab \RR,
 \qquad
 \Tens{C} = \nab \RR\cdot \(\nab\RR\)^T,
\ee
where $\Tens{C}$ is the Cauchy-Green deformation measure.

Equation of state for the thermal stresses is derived as follows. Consider the specific thermal energy
\be{}
 U_T = U  - U_0, \qquad  U_0 = U|_{\At_\alpha=0} = \frac{1}{2} \sum_\alpha \Pi\(A_\alpha\).
\ee
where~$U$ is defined by formula~\eq{Utdef}.
According to formula~\eq{Utdef}, the thermal energy has kinetic and potential parts. The kinetic part is represented
as a function of~$\At_\alpha$ using the virial theorem~\citep{Krivtsov 2011 MTT}:
\be{KT_virial 3D}
 \frac{m}{2}\av{\vt^2} \approx \frac{1}{4} \sum_{\alpha} \av{\At_\alpha \cdot \F_\alpha\({\bf A}_\alpha + \At_\alpha\)}.
\ee
\added{Then thermal stresses and thermal energy depend on parameters~$\At_\alpha$ characterizing thermal motion.}

In the first approximation, series expansion of $\cst_T$ and $U_T$ with respect to~$\At_\alpha$ yields
\be{pU1a}
\begin{array}{l}
    \DS \cst_T \approx -\frac{1}{2V}\sum_{\alpha} \left[
        2\Phi'{\bf A}_{\alpha} {\bf E} {\bf A}_{\alpha} +
        \Phi'{\bf A}_{\alpha}{\bf A}_{\alpha}{\bf E}+
        2\Phi''{\bf A}_{\alpha}{\bf A}_{\alpha}{\bf A}_{\alpha}{\bf A}_{\alpha} \right] \cdot\cdot
        \,\av{ \At_\alpha \At_\alpha },
 \\ [6mm]
    \DS U_T  \approx -\frac{1}{2} \sum_{\alpha} \left[    \Phi \Tens{E}+   2\Phi' \A_{\alpha}\A_{\alpha} \right] \cdot\cdot
    \,\av{ \At_\alpha \At_\alpha }.
 \end{array}
 \ee
Equations~\eq{pU1a} show that thermal energy and thermal pressure depend on symmetric second rank tensors~$\av{\At_\alpha\At_\alpha}$. In order to obtain the equation of state,  additional relations between components of these tensors are required.

In paper~\citep{Krivtsov 2011 MTT}, the following assumption is used
\be{assumption}
    \av{\At_\alpha\At_\alpha} =  \frac{1}{d}\,\xi^2\Tens{E},
     \qquad
     \xi^2 = \av{\At_\alpha^2}.
\ee
\added{In this case, thermal pressure and thermal energy are functions of a single scalar parameter~$\xi^2$ characterizing thermal motion.
Excluding this parameter from formulas~\eq{pU1a}, yields the equation of state}
\be{}
  \DS \cst_T =\frac{\Tens{G}}{V} U_T,
  \qquad
   \Tens{G}
   =
    \frac{\displaystyle
    \sum_{\alpha}\left( (d+2)\Phi' + 2\Phi'' A_\alpha^2 \right)
    \A_{\alpha}\A_{\alpha}}{\displaystyle
    \sum_{\alpha} \left( d\,\Phi + 2\Phi' A_\alpha^2 \right)},
 \end{equation}
where tensor~$\Tens{G}$ is related to conventional Gr\"uneisen parameter as~$\varGamma =  -\frac{1}{d} {\rm tr} \Tens{G}$.
In the case of interactions of the nearest-neighbors, the expression for the Gr\"uneisen parameter reads
\be{Gamma_simple}
 \varGamma
 =
 -\frac{\Pi'''A^2 + (d-1)\left[\Pi''A -\Pi'\right]}{2d(\Pi'' A + (d-1)\Pi')}.
\ee
A particular case of formula~\eq{Gamma_simple} for the face-centered cubic lattice~($d=3$)
is derived in paper~\citep{Stacey 1975}. Note that according to formula~\eq{Gamma_simple}, the Gr\"uneisen parameter can be negative. This case corresponds to negative thermal expansion~\citep{Kuzkin 2014 JPC}, \citep{Dove and Fang 2016}.

In one-dimensional case,  assumption~\eq{assumption} is satisfied, and therefore formula~\eq{Gamma_simple} is exact. In multidimensional case, computer simulations show that assumption~\eq{assumption} is not accurate --- tensors~$\av{\At_\alpha\At_\alpha}$ are not isotropic. Then
additional relations are needed to close the system of equations~\eq{pU1a}.

In two-dimensional case,
 equations~\eq{pU1a} contain additional unknown parameters
\be{}
 \beta_\alpha = \frac{\av{\(\At_\alpha \cdot\Vect{n}_\alpha\)^2}}{  \av{\(\At_\alpha\cdot\Vect{e}_\alpha\)^2}},
\ee
characterizing the relation between the longitudinal and \added{in-plane} transverse deformations of the bonds caused by the thermal motion. Here~$\Vect{e}_\alpha = \va_\alpha/|\va_\alpha|$; \added{$\Vect{n}_\alpha$ is normal to~$\Vect{e}_\alpha$ in the lattice plane.} For triangular lattice with nearest-neighbor interactions,~$\beta_\alpha$ is independent on~$\alpha$. Therefore index $\alpha$ is omitted below.
Then the Gr\"uneisen parameter is represented in terms of~$\beta$ as follows~\citep{Panchenko 2017 DAN}
\be{Gamma_2D}
 \varGamma
 =
 -\frac{\Pi'''A^2 + \beta\left[\Pi''A -\Pi'\right]}{4(\Pi'' A + \beta \Pi')},
\ee
Parameter~$\beta$ can be estimated using harmonic crystal model. In paper~\citep{Kuzkin 2017 PhysSolState}, an equation for the covariance of the particle displacements is derived. Numerical solution of this equation yields the value of parameter~$\beta \approx 1.43$. This value is in a good agreement with results of molecular dynamics simulations~\citep{Panchenko 2017 DAN}.

In three-dimensional case, for each bond there are two unknown parameters  characterizing the relation between longitudinal deformation and transverse deformations in two different directions. Molecular dynamics investigation of this problem is carried out in paper~\citep{Barton Stacey 1985}. Parameters similar to~$\beta_\alpha$ are calculated for the face-centered cubic lattice with Lennard-Jones interactions.

Thus series expansion of the stress tensor and the thermal energy allows to derive  equations of state. In the first approximation, the equation of state in generalized~(tensor) Mie-Gr\"uneisen form is obtained. Leaving more terms in the series, yields more accurate equations of state similar to the equations~\eq{EOS_ex2}~(see paper~\citep{Krivtsov 2011 MTT}). Note that for two- and three-dimensional crystals, nonlinear corrections to the Mie-Gr\"uneisen equation of state are less important than for one-dimensional or quasi-one-dimensional crystals.

\section{Heat Transfer In One-dimensional Crystal}

\subsection{Nonlocal Temperature}

As it is shown above, it is possible to rigorously derive  macroscopic continuum equations for the anharmonic crystals in the case of adiabatic approximation, where the heat fluxes are neglected. Attempts to obtain by the similar way the constitutive equation for the heat flux failed. In continuum mechanics, the Fourier law is widely used. The law assumes linear dependence between the heat flux and temperature gradient.
However, this law is not fulfilled for harmonic and weakly anharmonic crystals~\citep{Rieder, Lepri Livi Politi,Dhar altern mass,Harris et al 2008,Gendelman 2010, Dudnikova 2003}. As it is shown below, generally an infinite number of additional variables (generalized energies or nonlocal temperatures) should be added to obtain the closed equations for the heat transfer~\citep{Krivtsov 2014 DAN, Krivtsov 2015 DAN, Krivtsov 2015 arxive, Kuzkin 2017 arxive}.

Consider one-dimensional crystal \eq{4} in harmonic approximation
\be{h1}
     \ddot{u}_k = {\cal L}_k u_k \= \w_e^2(u_{k-1}-2u_k+u_{k+1})
     \qq \w_e \= \sqrt{C/m}
,\ee
where
 ${\cal L}_k$ is the linear difference operator applied to index $k$;
 coefficient $\w_e$ is the elementary frequency;
 $C \= \Pi''(a)$ is the bond stiffness.

The thermal energy and the kinetic temperature are not sufficient for description of the heat transfer. To close the system of equations the generalized nonlocal temperatures are introduced:
\be{h2}
    \kB\theta_{pq} \= m\av{\tilde v_p\tilde v_q}
.\ee
The nonlocal temperature satisfies the following differential-difference equation~\citep{Krivtsov 2006 APM}
\be{h3}
    \ddddot\theta_{\!\!\!pq} - 2({\cal L}_p+{\cal L}_q)\ddot\theta_{pq} + ({\cal L}_p-{\cal L}_q)^2\theta_{pq} = 0.
\ee
This equation describes two processes: fast transition to the local thermal equilibrium~\citep{Krivtsov 2014 DAN} and slow heat transfer~\citep{Krivtsov 2015 DAN, Krivtsov 2015 arxive}.

If only the slow motion is considered then the first term with the forth derivative in equation \eq{h3} can be neglected resulting in the equation of the second order with respect to time. For continualization new variables are introduced
\be{h4}
    \theta_k(x) \= (-1)^k\theta_{pq} \qq
    k\=q-p\qq x\=\frac{p+q}{2}\,a
,\ee
where $x$ is the macroscopic spacial coordinate.
Then the nonlocal temperature $\theta_k(x)$ satisfies equation~\citep{Krivtsov 2015 DAN, Krivtsov 2015 arxive}
\be{h5}\TS
    \ddot\theta_k + \frac14c^2(\theta_{k-1}-2\theta_{k}+\theta_{k+1})'' = 0,
\ee
where $c\=\w_e a$ is the sound velocity in the crystal. Equation \eq{h5} can be interpreted as an infinite system of coupled wave equations. Given known the solution of equation~\eq{h5}, the kinetic temperature is calculated as $T(x)=\theta_k(x)|_{k=0}$.

\subsection{Heat Impact}

The initial problem of the heat impact for equation \eq{1} is
\be{h6}
     u_k|_{t=0} = 0
     \qq
     \dot u_k|_{t=0} = \sigma(x)\varrho_k
,\ee
where $\rho_k$ are independent random values with zero expectation
and unit variance, $\sigma(x)$ is variance of the initial
velocities, which is a slowly varying function of the spatial
coordinate $x = k a$.
This kind of initial conditions can be induced by an ultrashort laser pulse~\citep{Inogamov et al}, \cite{Indeytsev et al}.

The corresponding initial conditions for the nonlocal temperature are
\be{h7}
    \theta_k(x)|_{t=0} = T_0(x)\delta_k \qq
    \dot\theta_k(x)|_{t=0} = 0 ,
\ee
where $T_0(x) = m\sigma^2(x)/(2\kB)$ is the initial temperature distribution,
${\delta_{k} = 1}$ for ${k=0}$ and ${\delta_{k} = 0}$ for $k\ne 0$.
These initial conditions correspond to the end of the fast transition process~\citep{Krivtsov 2014 DAN}, resulting in double decrease of the initial kinetic temperature due to equilibration between the kinetic and potential parts of the thermal energy (according to the virial theorem).

Solution of the initial problem \eq{h5}, \eq{h7} yields the following expression for the kinetic temperature~\citep{Krivtsov 2015 DAN, Krivtsov 2015 arxive}
\be{h8}
    T(t,x)
    = \frac1\pi\int_{-1}^{1}\frac{T_0(x-c t s)}{\sqrt{1-s^2}}\,{\rm d}s
    = \frac1{2\pi}\int_{0}^{2\pi}T_0(x+ct\cos{\TS\frac p2})\,{\rm d} p
.\ee
It can be shown that $c\cos{\TS\frac p2}$ is the functional dependence of the  group velocity of equation \eq{1} on the wave number $p$. Then the second solution in \eq{h8} can be interpreted as superposition of waves traveling with group velocity and having a shape of initial temperature distribution~\citep{Kuzkin 2017 arxive}.

Solutions \eq{h8} satisfy the following differential equation
\be{h9}
    \ddot T +\frac1t\,\dot T = c^2 T''
\ee
with initial conditions
\be{h10}
    T|_{t=0} = T_0(x) \qq
    \dot T|_{t=0} = 0
.\ee
Equation \eq{h9} is a particular case of the Darboux equation. Equation \eq{h9} for thermal processes in harmonic one-dimensional crystal is originally derived in~\citep{Krivtsov 2015 DAN, Krivtsov 2015 arxive}. This equation is non-autonomous: one of its coefficients explicitly depends on time~$t$. Despite the fact that the coefficient in equation~\eq{h9} has singularity for $t=0$, its solution with initial conditions \eq{h10}, as it follows from~\eq{h8}, has no singularities for any smooth $T_0(x)$. Equation~\eq{h9} is non-autonomous because it describes an evolution of the heat impact --- the sudden heat perturbation~\eq{h6} happened at $t=0$. The coefficient $t$ in the equation is the time elapsed from the moment of the heat impact. That is why equation \eq{h9} is not time-invariant (it changes with the time shift $t\to t+\tau$) and it can be considered only with initial conditions~\eq{h10}. General heat transfer processes are described by equation~\eq{h5}, which is much more complicated, but it has constant coefficients and therefore it is autonomous and time-invariant.

Equation~\eq{h9} looks similar to the equation of hyperbolic heat conductivity
\be{h11}
    \ddot T +\frac1\tau\,\dot T = c^2 T''
,\ee
where $\tau$ is the relaxation constant, $c$ is the wave front velocity. Indeed, both equations demonstrate wave behavior with the finite speed~$c$ for the front propagation. This differs them from the classic Fourier heat equation $\dot T = \beta T''$ ($\beta$ is the thermal diffusivity). For the Fourier equation a signal propagates with an infinite speed and therefore the heat front is absent. Equation \eq{h11} is empiric, while equation~\eq{h9} is rigorously derived from lattice dynamics equations~\eq{1}.


\subsection{Heat Flux}

The heat flux~\eq{8} in the harmonic case reads
\be{h12}
    h
    = -\av{{\widetilde F}_k{\widetilde v}_{k-1}}
    = -C\av{{\widetilde \epsilon}_k{\widetilde v}_{k-1}},
\ee
where $C$ is the bond stiffness.
Time differentiation with subsequent continualization yields
\be{h13}
    \dot h
    = -\frac12\,\rho c^2\kB(T-\theta_1)'
\qq
    \kB\theta_1 \= -\av{{\widetilde v}_k{\widetilde v}_{k-1}}
.\ee
Therefore, once the equation \eq{h5} for the nonlocal temperatures is solved, the heat flux can be obtained from relation \eq{h13}. However it is seen that the heat flux depends not only on the kinetic temperature $T$, but also on the nonlocal temperature~$\theta_1$. That is why it is impossible to close termomechanics equations in general nonadiabatic case without using equation \eq{h5} or its nonlinear extension.

For the heat impact problem, the solution of equation \eq{h5} yields the following constitutive relation for the heat flux~\citep{Krivtsov 2015 DAN, Krivtsov 2015 arxive}
\be{h14}
    \dot h +\frac1t\,h = -\rho c^2 \kB T'
,\ee
which is an analogue of the Fourier law for the considered system.

\section{Summary}

 Equations of thermomechanics for both discrete and continuum levels consist of balance equations (balance of momentum, energy and etc.) and constitutive equations (equations of state). At the discrete level, the constitutive equations relate the bond deformations and forces; at the continuum level --- the temperature (or thermal energy), strains, stresses, heat fluxes, etc.
 Balance equations can be obtained rigorously at both levels. Derivation of constitutive equations is much more complicated. Fundamental laws of thermodynamics and the principle of material objectivity yields some restrictions on the structure of constitutive equations. However the ambiguity in the construction of constitutive equations is rather large. In this situation, it is useful to consider discrete systems, for which the constitutive equations can be derived analytically.

 Transition from discrete to continuum description is demonstrated above for a number of relatively simple, but still challenging systems. In the adiabatic case~(zero heat flux), the transition can be carried out for anharmonic crystals. Series expansion of stress and internal energy with respect to small parameter characterizing thermal motion yields equations of state. In two- and three-dimensional cases, the equation of state in the Mie-Gr\"uneisen form has sufficient accuracy. For one-dimensional and quasi-one-dimensional crystals the Mie-Gr\"uneisen equation of state can be inaccurate or even qualitatively wrong. In this case, more accurate nonlinear equations of state described above should be used.

If the heat transfer is taken into account, the situation is more complicated. Therefore only the simplest harmonic system was considered. Generally, the presented approach for description of heat transfer can be extended to arbitrary harmonic systems and some anharmonic systems, but this requires additional considerations. This work was supported by the Russian Science Foundation~(RSCF grant No. 17-71-10213).



\end{document}